\documentclass[12pt]{article}
\usepackage{amsmath,amsfonts,amssymb,mathrsfs,cite}
\usepackage{color,graphicx,shortvrb,epsfig}
\usepackage{latexsym}
\usepackage[a4paper]{geometry}
\usepackage[in]{fullpage}
\newlength{\myem}
\newcounter{mysubequation}[equation]

\makeatletter
\renewcommand{\section}{\@startsection{section}{1}{0em}{-\baselineskip}%
{\baselineskip}{\normalfont\large\bfseries}}
\renewcommand{\subsection}%
{\@startsection{subsection}{2}{0em}{-0.7\baselineskip}%
{0.7\baselineskip}{\normalfont\bfseries}}
\makeatother

\newcommand{\ba}{\begin{array}}
\newcommand{\ea}{\end{array}}
\newcommand{\bd}{\begin{displaymath}}
\newcommand{\ed}{\end{displaymath}}
\newcommand{\bi}{\begin{itemize}}
\newcommand{\ei}{\end{itemize}}
\newcommand{\benu}{\begin{enumerate}}
\newcommand{\eenu}{\end{enumerate}}
\newcommand{\be}{\begin{equation}}
\newcommand{\ee}{\end{equation}}
\newcommand{\bea}{\begin{eqnarray}}
\newcommand{\eea}{\end{eqnarray}}
\newcommand{\HKd}{\mbox{\sf HK}}

\newcommand{\UNOd}{\mbox{\sf UNO}}
\newcommand{\INOd}{\mbox{\sf INO}}
\newcommand{\MEMPHYSd}{\mbox{\sf MEMPHYS}}

\newcommand{\ie}{{\it i.e.}}

\newcommand{\eg}{{\it e.g.}}

\newcommand{\eq}{Eq.}

\def\dji{\mathrm{\Delta m_{ji}^2}}
\def\da{\mathrm{\Delta m_{31}^2}}

\def\ds{\mathrm{\Delta m_{21}^2}}
\def\pee{{{\rm P_{e e}}}}
\def\pmue{{{\rm P_{\mu e}}}}
\def\pemu{{{\rm P_{e \mu}}}}
\def\pmumu{{{\rm P_{\mu \mu}}}}
\def\signda{{{{\sf{sign}}}}{(\Delta {\rm{m_{31}^2}})}}
\newcommand{\chr}{\mbox{$\breve{\rm C}$erenkov~}}
\begin{document}
%
%
% ~~~~~~~~~~~~~~~~~~~~~~~~~~~~~~~~~~~~~~~~~~~~~~~~~~~~~~~~~~~~~~~~~                    Title-page
%
%
\begin{flushright}
%{\makebox[1.cm]
%        \sf hep$\,$-$\,$ph/0705xxx}
\end{flushright}
\vspace*{1cm}
\setcounter{footnote}{-1}
    {\begin{center}
    {\LARGE{\bf{Resolving the Mass Hierarchy  
with Atmospheric Neutrinos using a Liquid Argon Detector}}}
    \end{center}}
\renewcommand{\thefootnote}{\fnsymbol{footnote}}
\vspace*{1cm}
                {\begin{center}
                {{\bf
                Raj Gandhi $^{a, \,\!\!\!}$
                \footnote[1]{\makebox[0.cm]{}
                \sf raj@mri.ernet.in},
                Pomita Ghoshal $^{b, \,\!\!\!}$
                \footnote[2]{\makebox[0.cm]{}
                \sf pomita@theory.tifr.res.in},
                Srubabati Goswami $^{a,c, \,\!\!\!}$
                \footnote[3]{\makebox[0.cm]{}
                \sf sruba@prl.res.in},
                S Uma Sankar $^{d, \,\!\!}$
                \footnote[5]{\makebox[0.cm]{}
                \sf uma@phy.iitb.ac.in}
                }}
                \end{center}}
\vskip 1.2cm
{\small
                \begin{center}
                $^a$ Harish-Chandra Research Institute, Chhatnag Road, \\
                     Jhunsi, Allahabad 211 019, India\\[4mm]
                $^b$ Department of Theoretical Physics,Tata Institute of Fundamental Research, \\
                     Colaba, Mumbai 400 005, India\\[4mm]
                $^c$ Physical Research Laboratory, \\
                     Navrangpura, Ahmedabad 380 009, India\\[4mm]
                $^d$ Department of Physics, Indian Institute of Technology, 
Powai,\\
                     Mumbai 400 076, India
                \end{center}}

\vspace*{0.5cm}
\date{today}
%
%
% ~~~~~~~~~~~~~~~~~~~~~~~~~~~~~~~~~~~~~~~~~~~~~~~~~~~~~~~~~~~~~~~~~                 Abstract
%
%
\begin{abstract}
We explore the potential offered by large-mass Liquid Argon
detectors for determination of the sign of $\da$, or the neutrino
mass hierarchy, through interactions of atmospheric neutrinos. We
give results for a  100 kT sized magnetized detector which provides separate
sensitivity to $\nu_\mu$ , $\bar{\nu}_\mu$ and, over a limited
energy range, to $\nu_e$ , $\bar{\nu}_e$ . 
We also discuss the sensitivity for the unmagnetized version
of such a detector.
After including the
effect of smearing in neutrino energy and direction and
incorporating the relevant statistical,
theoretical and systematic errors, we perform a binned $\chi^2$
analysis of simulated data. The $\chi^2$ is marginalized over the
presently allowed ranges of neutrino parameters and determined as a
function of $\theta_{13}$. We find that such a detector offers
superior capabilities for hierarchy resolution, allowing a $> 4\sigma$
determination for a 100 kT detector 
over a 10 year running period for values of $\sin^2
2\theta_{13}  \ge 0.05$. For an unmagnetized detector,
a $2.5\sigma$ hierarchy sensitivity is possible for 
$\sin^2
2\theta_{13} = 0.04$.
\end{abstract}
%
%
% ~~~~~~~~~~~~~~~~~~~~~~~~~~~~~~~~~~~~~~~~~~~~~~~~~~~~~~~~~~~~~~~~~                    Titlepage ends
%

\newpage
\renewcommand{\thefootnote}{\arabic{footnote}} % To put the correct numbers for footnotes
\setcounter{footnote}{0}
%
% ~~~~~~~~~~~~~~~~~~~~~~~~~~~~~~~~~~~~~~~~~~~~~~~~~~~~~~~~~~~~~~~~~                    Introduction
%
%
\section{Introduction}
Parameters related to neutrino masses and mixings play a fundamental
role in  efforts to construct a viable theory beyond the
Standard Model. Over the last decade, our knowledge of these
parameters has increased at an unprecedented pace, due to crucial
results from solar, atmospheric, reactor and accelerator
based neutrino oscillation experiments. (For a recent review, we
refer the reader to ~\cite{Gonzalez-Garcia:2007ib} and references therein.)

Among the important but as yet unanswered questions, one of the
foremost is the determination of $\signda$ or the  hierarchy of
neutrino masses\footnote{We use the convention ${\mathrm{\dji \equiv
m_j^2 - m_i^2. }}$}, presently unconstrained by available data. If
$\signda
> 0$, then we have the mass pattern, ${\mathrm{m_3 \gg m_2 \gg
m_1}}$, which is similar to that of  the charged leptons. This is
called the normal hierarchy (NH). If $\signda < 0$, then the mass
pattern is ${\mathrm{m_2 \geq m_1 \gg m_3}}$. This is called the
inverted hierarchy (IH). In recent papers, the prospects for
progress towards a resolution of this question offered by
atmospheric neutrinos have been
explored~\cite{Banuls:2001zn,Bernabeu:2001xn,Indumathi:2004kd,Gandhi:2004bj,
Petcov:2005rv,Bernabeu:2003yp,PalomaresRuiz:2004tk,Samanta:2006sj,Gandhi:2007td}. With most of their flux
 below 10 GeV, these neutrinos traverse distances up to $\sim$ 12000 km in the earth's matter enroute to a detector.
 This exposes them to  appreciable  resonant matter effects which occur (primarily) for  energies 
between 2$-$10 GeV and  distances between
4000$-$12500 km inside the earth~\cite{Gandhi:2004md,Gandhi:2004bj,Akhmedov:2006hb}.
Until very long baseline experiments using $\beta$-beams or neutrino
factories~\cite{Group:2007kx} are
built,  atmospheric neutrinos  permit us to exploit these effects,
albeit in a slower and less spectacular fashion which calls for a
careful analysis of accumulated effects over many  bins in energy
and angle. Such an approach was recently pursued
in~\cite{Petcov:2005rv,Gandhi:2007td} for megaton water \chr detectors
(\eg~\HKd~\cite{Nakamura:2000tp,Itow:2001ee},
\UNOd~\cite{Jung:1999jq} or \MEMPHYSd~\cite{deBellefon:2006vq}) and
magnetized iron detectors (\eg~\INOd~\cite{Athar:2006yb}). 
The salient features emerging out of the above analyses are 
\begin{itemize}
\item Both muon and electron events arising from atmospheric 
neutrinos have sensitivity to sign of $\Delta m^2_{31}$ for 
sufficiently large values of $\sin^2 2\theta_{13}$ . 
\item One of the major factors responsible for 
a reduced sensitivity is the  finite energy and angular smearing of a detector.
\item The muon events are more sensitive to smearing than electron events. 
\item Magnetized iron calorimeter detectors have the advantage of  
having charge identification capability but they are sensitive only to 
muon events 
\item Water \chr detectors have the advantage that they are sensitive to both 
muon and electron events. On the other hand they do not have charge sensitivity. 
\end{itemize} 
The above features indicate that a detector which is sensitive both to 
muon and electron events as well as their charges will be ideal for 
probing hierarchy of neutrino masses. 
An important class of
future detectors using Liquid Argon as their active
medium, may provide this kind of set up specially if 
there can be magnetized versions of these. 
In this
paper we study the hierarchy sensitivity of Liquid Argon detectors 
for cases with and without charge sensitivity.

%extend this approach to include
% an important class of
%future detectors, \ie those using Liquid Argon as their active
%medium.

%
\section{Basic Characteristics of Liquid Argon Detectors}
Liquid Argon detectors are time projection chambers with fine-grained
tracking and total absorption calorimetry. Ionization electrons
resulting from the passage of an energetic particle through the
medium are detected by drifting their paths over several meters to
wire planes. The orientation of these planes is designed to
reconstruct the time, length and position of each path by recording
multiple snapshots of the  electrons, from which a bubble-chamber
like image is constructed. The viability of this technology for a
small detector has been convincingly demonstrated
by~\cite{Ankowski:2006ts}, and intensive efforts are underway for
the upgraded development of large mass detectors~\cite{
Cline:2006st,Rubbia:2004tz,Ereditato:2004ru} to fully exploit the
promise of this technique.

Essentially, the technique allows us to detect charged particles
with good resolution over the range of MeV to multi GeV.  It uses
well studied $dE/dx$ measurements in the medium to separate
electrons, muons, pions, kaons and protons from each other. It is
also possible to separate the light from electrons vs that from
neutral pions with high efficiency. Magnetization over a 100 kT
volume has been deemed
possible~\cite{Ereditato:2005yx,Badertscher:2005te}.
As mentioned earlier, we present results for both the magnetized
and unmagnetized case.

% ({\bf
%\it {Note: {It would be useful to present results both with and
%without a magnetic field. Experimentally, providing the field over
%such a large volume of argon may be a nontrivial problem}}})

We assume the following
energy resolutions over the ranges that will be relevant to our
calculations \cite{Rubbia:2004tz}: \\
For the GeV electrons that we will be interested in,
\be
\sigma_{e}/E_e = 3\% /{\sqrt E_e}  
\ee 
where $E_e$ is the electron energy and $\sigma_{e}$ is the energy resolution 
of the electron. All values of energy are expressed in GeV.
For the muon neutrino we take the
energy resolution in terms of the muon energy to be
\be
\sigma_{\mu}/E_{\mu} = 15\%
\ee

Additionally, for hadronic
showers , which account for a significant fraction of the
uncertainty in the determination of the primary neutrino energy, we
assume
\be
\sigma_{had}/E_{had} = 30\% /{\sqrt E_{had}} 
\ee 
where $E_{had}$ is the hadron energy and $\sigma_{had}$ is the energy resolution
of the hadron.

Our procedure for
inferring the neutrino energy from the measured charged lepton and
hadronic energies incorporates a knowledge of the average rapidity
for charged current cross-sections in this energy region.
The energy resolution in terms of the neutrino energy is related to the leptonic and hadronic
energy resolutions as follows:
\be
\sigma_{\nu}/E_{\nu} = {\sqrt{(1-y)^2 (\sigma_{lep} / E_{lep})^2 + y^2 (\sigma_{had}/E_{had})^2}}
\ee
The rapidity is defined as $y = E_{had}/E_{\nu}$, where $E_{\nu} = E_{lep} + E_{had}$ is the energy of the neutrino.
Therefore, the energy resolution in terms of the neutrino energy is given by
\be
\sigma_{\nu_e}/E_{\nu_e} = {\sqrt{(1-y)(0.03)^2 / E_{\nu_e} + y (0.3)^2/E_{\nu_e}}}
\ee
and
\be
\sigma_{\nu_{\mu}}/E_{\nu_{\mu}} = {\sqrt{(1-y)^2 (0.15)^2  + y (0.3)^2/E_{\nu_{\mu}}}}
\ee
for electron and muon neutrinos respectively.

In our computation, we take the average rapidity in the energy region of our interest (i.e. in the GeV range) 
to be 0.45 for neutrinos and 0.3 for antineutrinos \cite{Gandhi:1995tf}. 
%For the muon neutrino we take the 
%energy resolution in terms of the muon neutrino energy to be
%$$
%\sigma/E_{\nu} = 15\%
%$$
The angular resolution of the detector is taken to be $\sigma_{\theta} = 10^o$. 
The energy threshold and ranges in which charge identification is feasible are 
$E_{threshold}  = 800$ MeV for charge identification of muons with high efficiency and
$E_{electron}  = 1-5$ GeV  for charge identification of electrons with high efficiency.
Charged lepton detection and separation (e vs $\mu$) without charge identification 
is possible for $E_{lepton}>$ few MeV.
Also, for electron events, it is expected to have a 20$\%$ probability of $\sim 100\%$ charge identification in the
energy range 1 - 5 GeV.

The next section provides a brief description of our calculational
and numerical procedure, prior to our discussion of the results.

\section{Calculational and Numerical Procedure}

We follow the procedure described in~\cite{Gandhi:2007td}, and refer
the reader to the  description provided there for any details that
may be omitted here.

Our calculation of the  atmospheric electron and muon event rates
uses the neutrino oscillation probabilities corresponding to the
disappearance channels $\pmumu$ and  $\pee$ and appearance channels
$\pmue$ and $\pemu$ for both neutrinos and
antineutrinos\footnote{${\mathrm{P_{\alpha\beta}}}$ denotes the
probability for transition from $\nu_{\alpha} \to \nu_{\beta}$. }.
We have modeled the density profile of the earth by the Preliminary
Reference Earth Model (PREM)~\cite{prem} in order to  numerically
solve the full three flavour neutrino propagation equation through
the earth. The total CC cross section  used here is the sum of
quasi-elastic, single meson production and deep inelastic cross
sections~\cite{Ashie:2005ik,Huber:2004ka,SK1,Paschos:2001np}. For
the incident atmospheric neutrino fluxes we use the tables
from~\cite{Honda:2004yz}.
%We take into account the smearing in both
%energy and zenith angle, assuming Gaussian  resolution, and, for the
%angular part, including both polar and azimuthal smearing.

For our analysis in Liquid Argon detectors, we consider an exposure of 1 Mt yr (100 kT $\times$ 10 years).
We look at
the neutrino energy range of 1 - 10 GeV and the cosine of the zenith
angle ($\theta$) range of -1.0 to -0.1. These ranges are divided into 
9 bins in energy and 18 bins in zenith angle. The $\mu^-$
event rate in a specific energy bin with width $dE$ and the solid
angle bin with width ${\mathrm{d \Omega}}$ is expressed as :
\begin{equation}
\rm{ \frac{d^2 N_{\mu}}{d \Omega dE} = \frac{1}{2\pi} ~\left[
\left(\frac{d^2 \Phi_\mu}{d \cos \theta dE}\right) P_{\mu\mu} +
\left(\frac{d^2 \Phi_e}{d \cos \theta dE}\right) P_{e\mu}\right]
~\sigma_{CC} ~D_{eff} } \label{muevent}
\end{equation}
Here $\rm \Phi_{\mu,e}$ are the atmospheric fluxes ($\nu_\mu$ and
$\nu_e$), $\rm{\sigma_{CC}}$ is the total muon-nucleon charged
current cross-section and $\rm{D_{eff}}$ is the detector efficiency.
The $\mu^+$ event rate is similar to the above expression with the
fluxes, probabilities and cross sections replaced by those for
anti-muons. Similarly, the $e^-$ event rate in a specific energy and
zenith angle bin is expressed as follows:
\begin{equation}
\mathrm{ \frac{d^2 N_e}{d \Omega dE} = \frac{1}{2\pi} ~\left[\left(
\frac{d^2 \Phi_{\mu}}{d \cos \theta dE}\right) P_{\mu e} +
\left(\frac{d^2 \Phi_e}{d \cos \theta dE}\right) P_{ee} \right]
~\sigma_{CC} ~D_{eff} } \label{eevent}
\end{equation}
with the $e^+$ event rate being expressed in terms of anti-neutrino
fluxes, probabilities and cross sections.
E and $\theta$ in the above equations are true values
of neutrino energy and zenith angle.
We convert the above double differential event rates
into those with respect to the measured neutrino energy ${\rm{E_m}}$
and the measured zenith angle $\theta_m$.
This is done using Gaussian resolution functions
in energy and zenith angle, as explained in \cite{Gandhi:2007td}.

In the limit when only statistical errors are taken into account,
the standard Gaussian definition of binned $\chi^2$ is:
\begin{equation}
{\mathrm{ \chi^2_{stat} = \sum_{i=E_m bins} \ \
 \sum_{j=\cos \theta_m
bins} ~\frac{ \left[~N_{ij}^{ex}-N_{ij}^{th}~\right]^2}
{N_{ij}^{ex}} }} \label{chisqstat}
\end{equation}
Here, ${\mathrm N_{ij}^{ex}}$ is the experimental and ${\mathrm
N_{ij}^{th}}$ is theoretical number of events in the
${\mathrm{ij^{th}}}$ bin.

However, in addition to the statistical uncertainties, one also
needs to take into account various theoretical and systematic
uncertainties. In particular, our analysis includes a flux
normalization error of 20$\%$, a tilt
factor~\cite{Gonzalez-Garcia:2004wg} which takes into account the
deviation of the atmospheric fluxes from a power law, a zenith angle
dependence uncertainty of 5 $\%$, an overall cross section uncertainty of 10 $\%$, 
and an overall systematic uncertainty of 5 $\%$ \cite{Gandhi:2007td}. 
These uncertainties are included
using the method of pulls described
in~\cite{Fogli:2002pt,Fogli:2003th,Gonzalez-Garcia:2004wg}.

In this method, the uncertainty in fluxes and cross sections and the
systematic uncertainties are taken into account by allowing these
inputs to deviate from their standard values in the computation of
${\mathrm{ N^{th}_{ij} }}$. Let the ${\mathrm{ k^{th} }}$ input
deviate from its standard value by ${\mathrm{ \sigma_k \;\xi_k }}$,
where ${\mathrm{ \sigma_k }}$ is its uncertainty. Then the value of
${\mathrm{ N^{th}_{ij} }}$ with the changed inputs is given by
\begin{equation}
{\mathrm{ N^{th}_{ij} =  N^{th}_{ij}(std) + \sum^{npull}_{k=1}\;
c_{ij}^k \;\xi_k }} \label{cij}
\end{equation}
where ${\mathrm{ N^{th}_{ij}(std) }}$ is the theoretical rate for
bin ${\mathrm{ij}}$, calculated with the standard values of the
inputs and npull is the number of sources of uncertainty, which in
our is case is 5.
The ${\mathrm{ \xi_k }}$'s are called the ``pull" variables and they
determine the number of ${\mathrm{ \sigma}}$'s by which the
${\mathrm{ k^{th} }}$ input deviates from its standard value. In
\eq~(\ref{cij}), ${\mathrm{ c_{ij}^k }}$ is the change in ${\mathrm{
N^{th}_{ij} }}$ when the ${\mathrm{ k^{th} }}$ input is changed by
${\mathrm{ \sigma_k }}$ (\ie~by 1 standard deviation). The
uncertainties in the inputs are not very large. Therefore, in
\eq~(\ref{cij}) we only considered the changes in ${\mathrm{
N^{th}_{ij} }}$ which are linear in ${\mathrm{ \xi_k }}$. Thus we
have a modified $\chi^2$ defined by
\begin{equation}
\mathrm{ {\chi^2(\xi_k)} = \sum_{i,j}\;
\frac{\left[~N_{ij}^{th}(std) \;+\; \sum^{npull}_{k=1}\; c_{ij}^k\;
\xi_k - N_{ij}^{ex}~\right]^2}{N_{ij}^{ex}} + \sum^{npull}_{k=1}\;
\xi_k^2  } \label{chisqxik}
\end{equation}
where the additional term ${\rm{\xi_k^2}}$ is the penalty imposed
for moving ${\mathrm{k^{th}}}$ input away from its standard value by
${\rm{\sigma_k \;\xi_k}}$. The $\chi^2$ with pulls, which includes
the effects of all theoretical and systematic uncertainties, is
obtained by minimizing ${\rm{\chi^2(\xi_k)}}$, given in
\eq~(\ref{chisqxik}), with respect to all the pulls ${\rm{\xi_k}}$:
\begin{equation}
{\mathrm{ \chi^2_{pull} = Min_{\xi_k}~ \left[~ \chi^2(\xi_k)~\right]
}} \label{chisqpull}
\end{equation}

However,  in general, the values of the mass-squared differences
$|\da|$ and $\ds$ and the mixing angles $\theta_{12}$, $\theta_{23}$
and $\theta_{13}$ can vary over a range that reflects the
uncertainty in our knowledge. To take into account the uncertainties
in these parameters, we define the {\bf{marginalized}} $\chi^2$ for
hierarchy sensitivity as,
\begin{eqnarray}
{\mathrm{ \chi^2_{min}}}  &=& {\mathrm{ Min \left[~ \chi^2 (\xi_k)
~+~ \left(\; \frac{|\da|^{true} - |\da|}{\sigma\;(|\da|)}
\;\right)^2 \right.}}
\nonumber \\
&&{\mathrm{ \left. ~+~ 
%\left(\; \frac{\sin^2 2\theta_{23}^{true} -
%\sin^2 2\theta_{23}}{\sigma\;(\sin^2 2 \theta_{23})} \;\right)^2 ~+~
+ \left(\; \frac{\sin^2 2\theta_{13}^{true} - \sin^2 2
\theta_{13}}{\sigma\;(\sin^2 2\theta_{13})}\;\right)^2 ~\right] }}
\label{chisqfinal}
\end{eqnarray}
${\rm{\chi^2(\xi_k)}}$ in the above equation, is computed according
to the definition given in \eq~(\ref{chisqxik}).
The following values for the errors are chosen 
\begin{itemize}
\item
$\sigma\;(\sin^2 2\theta_{13}) = 0.02$,
\item
$\sigma\;(|\da|) = 10 \%~{\rm{of}}~|\da| = 0.25 \times 10^{-3} \ {\rm eV}^2$
\end{itemize}
In computing $\chi^2_{\rm min}$, we varied the values of the neutrino
parameters from their true values over $\pm 3 \sigma$ ranges. 
Note that we have not done the marginalization over the mixing
angle $\theta_{23}$. In this work, we have considered only one
value of $\theta_{23}^{\rm true} = \pi/4$. For this value of 
$\theta_{23}^{\rm true}$, it is found that the $\chi^2_{\rm min}$
always occurs at the true value itself.
We do not marginalize over the solar parameters $\theta_{12}$ and $\Delta m_{21}^2$ in our computation, 
since they are known with very good precision \cite{Bandyopadhyay:2008va}.

We have computed the values of the $\chi^2$ sensitivity to the mass hierarchy, choosing the true hierarchy to be normal.
The true values of the neutrino parameters are taken to be $\theta_{23} = 45^\circ$, $\theta_{12} = 33.8^\circ$,
$\vert \Delta m_{31}^2 \vert = 2.5 \times 10^{-3}$ eV$^2$, and 
$\Delta m_{21}^2  = 8.0 \times 10^{-5}$ eV$^2$
from the current $3\sigma$ allowed ranges \cite{Bandyopadhyay:2008va,:2008ee,Maltoni:2004ei,Aharmim:2008kc}.

\section{Results and Discussion}

We give our results for
an exposure of 1 Mt yr (unless otherwise stated) 
for two input values of $\theta_{13}$ and for detectors 
both with and without charge identification capability. 
When the detector has no charge identification capacity the 
$\chi^2$ is defined as, 
\be
(\chi^2_{m})_{noID} = \chi^2_{\mu + \bar{\mu}}
\ee
and
\be
(\chi^2_{el})_{noID} = \chi^2_{e + \bar{e}}.
\ee
That is, we sum over the particle and antiparticle events
and then compute the $\chi^2$. 
The total $\chi^2_{tot}$ is the sum of $\chi^2_m$ and $\chi^2_{el}$.

When the charge identification capability
is there, 
the $\chi^2$ values for muon events are computed as
\be
\chi^2_{m} = \chi^2_{\mu} + \chi^2_{\bar{\mu}}
\ee
where the individual $\chi^2$ for muon neutrino and 
muon antineutrino events are evaluated and then summed. 
We assumed that there is $100\%$ charge identification 
for muons throughout the energy range 1 - 10 GeV.
That, however, is not true for electrons. In fact,
there is no charge identification capability for electrons in 
the energy range 5 - 10 GeV and only a partial charge identification
capability for lower energy electrons. Therefore the 
$\chi^2$ for electron events is defined as 
\be
\chi^2_{el} = n (\chi^2_{e} + \chi^2_{\bar{e}})_{1-5 GeV} + 
(1-n) (\chi^2_{e+{\bar{e}}})_{1-5 GeV}
+ (\chi^2_{e+{\bar{e}}})_{5-10 GeV}.
\ee
Here, $n$ is the fraction of electron events in the energy range
1 - 5 GeV which are assumed to have exact charge identification capability.
As mentioned earlier, 
we expect a $\sim 20\%$ probability of exact charge identification in the
energy range 1 - 5 GeV. Hence we take $n=0.2$ in our 
computation unless otherwise mentioned. 
In other words, $20\%$ of the electron events
in this energy range are assumed to have $100\%$ charge sensitivity, 
while for the remaining $80\%$, the electron neutrino and 
antineutrino events are assumed to be indistinguishable. 
Realistically one expects a non-zero charge identification efficiency 
for the remaining $80\%$ electron events in this range, 
which would give a better sensitivity to the hierarchy. Hence 
we have given a conservative estimate in our computation 
of the hierarchy sensitivity of a Liquid Argon detector. 
We also give an estimate of how the sensitivity would vary if $n$ is varied
from 0 to 1.
The third term in the above expression is the 
$\chi^2$ contribution from the electron events in the 5 energy bins 
in the range 5 - 10 GeV, for which charge identification 
is not possible. Hence this $\chi^2$ value is
calculated using the sum of electron neutrino 
and antineutrino events in these energy bins.
$\chi^2_{el}$ represents the sum of all $\chi^2$ contributions 
from electron events. 

%Table \ref{fixnoID} gives the values of fixed-parameter $\chi^2_{m}$ and $\chi^2_{el}$ with and without pulls for different values of $\theta_{13}$ if we assume {\bf{no charge identification capability}} of a Liquid Argon detector.
%In this Table, 
%$(\chi^2_{tot})_{noID} = (\chi^2_{m})_{noID}+(\chi^2_{el})_{noID}$.

\begin{figure}[t]
   \centerline{\includegraphics[width=6in]{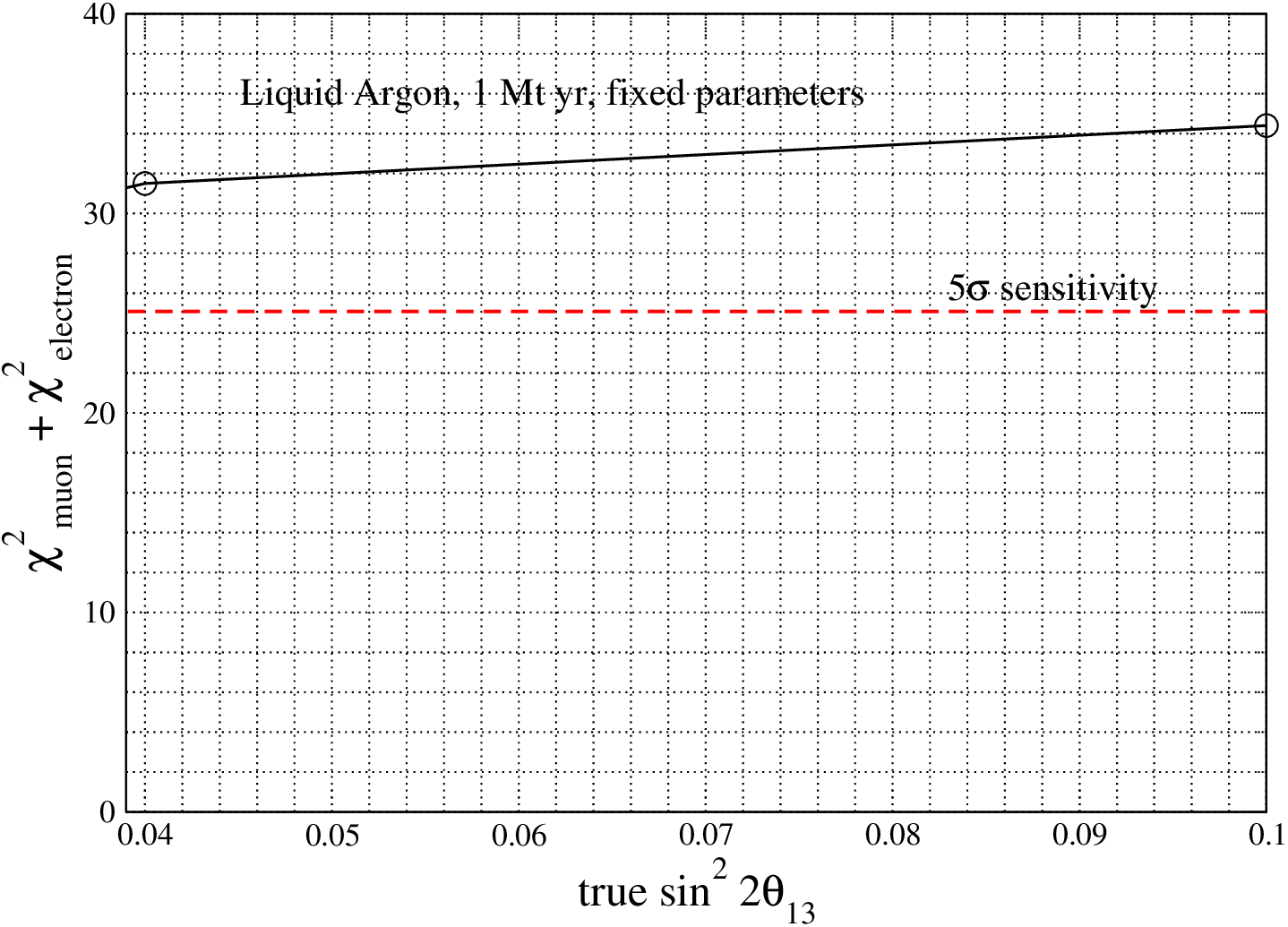}}
    \caption{Values of fixed parameter $\chi^2$ with pull versus
the input ({\it true}) value of $\sin^2 2 \theta_{13}$ for a Liquid
Argon detector (1 Mt yr). Shown is $(\chi^2_{tot})^{pull}$.}
%In these calculations,
%
%
%An energy resolution of
%$15 \%$ and an angular resolution of $10^\circ$ are assumed.
%Marginalization over neutrino parameters is done.
% The neutrino parameters used are $\stsmall = 0.1$ and $|\da| = + 2.5 \times 10^{-3} {\mathrm{eV^2}}$.
%}
\label{fig1}
\end{figure}

In Table \ref{fix}  we present the values of $\chi^2_m$ and $\chi^2_{el}$ with fixed
parameters for two different values of the parameter $\theta_{13}$. 
Here, $\chi^2_{tot} = \chi^2_m + \chi^2_{el}$ is the total 
fixed-parameter hierarchy sensitivity of a Liquid Argon detector with charge sensitivity.
In each case, $(\chi^2)^{stat}$ denotes the sensitivity with only 
statistical errors (\eq~(\ref{chisqstat}))
and  $(\chi^2)^{pull}$ denotes the sensitivity with theoretical 
and systematic uncertainties
incorporated as pulls (\eq~(\ref{chisqpull})).
The values of $(\chi^2_{tot})^{pull}$ are plotted against 
the input values of $\sin^2 2\theta_{13}$
in Figure \ref{fig1}.

\begin{figure}[t]
   \centerline{\includegraphics[width=6in]{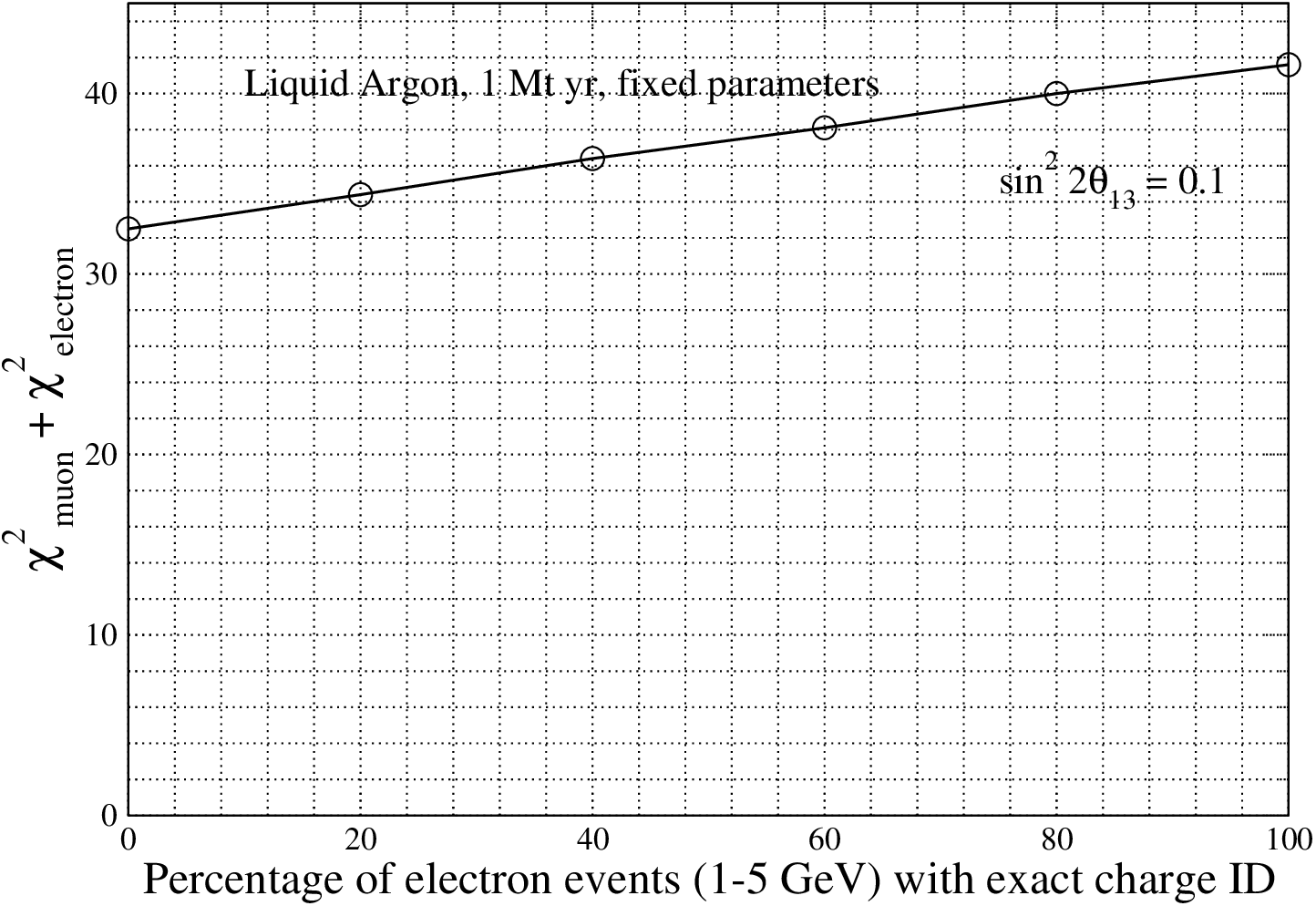}}
          \caption{Values of fixed parameter $\chi^2$ with pull versus
	  the percentage of electron events in the range 1-5 GeV taken to be
	  with 100 $\%$ charge ID efficiency for a Liquid Argon detector (1 Mt
	  yr). Shown is $(\chi^2_{tot})^{pull}$. Here,
	  $\sin^2 2\theta_{13} = 0.1$. }
	  %In these calculations,
	  %
	  %
	  %An energy resolution of
	  %$15 \%$ and an angular resolution of $10^\circ$ are assumed.
	  %Marginalization over neutrino parameters is done.
	  % The neutrino parameters used are $\stsmall = 0.1$ and $|\da| = + 2.5 \times 10^{-3} {\mathrm{eV^2}}$.
	  %}
	  \label{fig2}
	  \end{figure}

Figure \ref{fig2} shows the variation of  $(\chi^2_{tot})^{pull}$ (fixed-parameter) 
as $n$ varies from 0 to 1. 
%We observe that the dependence of the total $\chi^2$ on the charge identification 
%capability for electron events is not drastic, since the contribution 
%from the electron events in this energy range represent only a part of the total hierarchy 
%sensitivity of a Liquid Argon detector. 

\begin{figure}[t]
   \centerline{\includegraphics[width=6in]{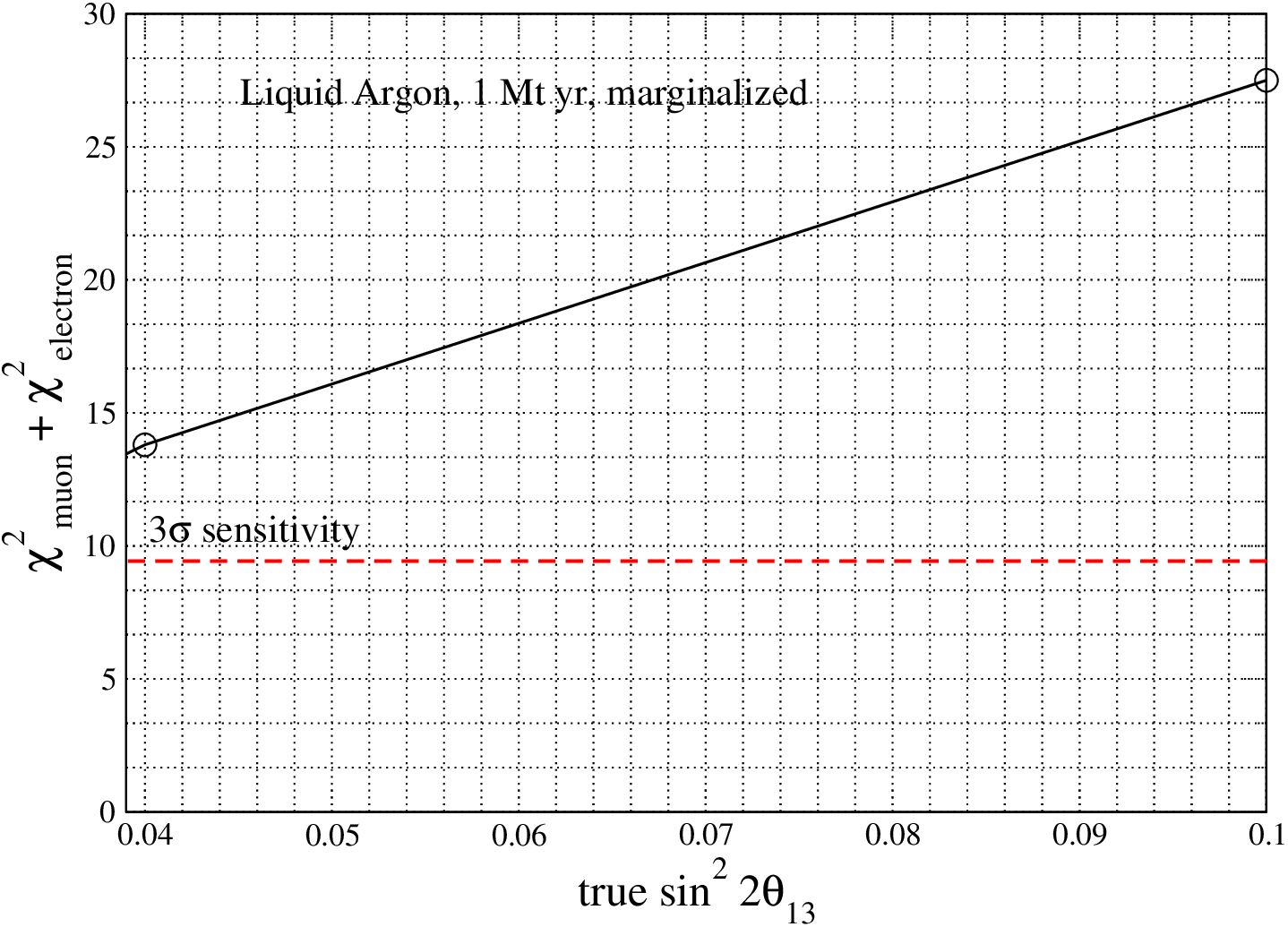}}
    \caption{Values of marginalized $\chi^2$ with pull and priors versus
the input ({\it true}) value of $\sin^2 2 \theta_{13}$ for a Liquid
Argon detector (1 Mt yr). Shown is $(\chi^2_{tot})^{pull}_{min}$}.
%In these calculations,
%
%
%An energy resolution of
%$15 \%$ and an angular resolution of $10^\circ$ are assumed.
%Marginalization over neutrino parameters is done.
% The neutrino parameters used are $\stsmall = 0.1$ and $|\da| = + 2.5 \times 10^{-3} {\mathrm{eV^2}}$.
%}
\label{fig3}
\end{figure}

In Table \ref{margin}  we give the total values of the $\chi^2$ hierarchy sensitivity with pulls and
priors with marginalization over the neutrino parameters $\theta_{13}$ and 
$\Delta m_{31}^2$ 
for two different values of $\theta_{13}^{true}$.
These values, denoted by $(\chi^2_{tot})^{pull}_{min}$, are for a Liquid Argon
detector with charge sensitivity and with an exposure of 1 Mt yr, and represent the principal results of this study. 
The same values are plotted against the input values of $\sin^2 2\theta_{13}$
in Figure \ref{fig3}.
For comparison, Table \ref{noID} gives the marginalized results for hierarchy sensitivity
for a Liquid Argon detector without charge sensitivity.

% calculated using
%Eq.\ref{chisq}. (Eq. for $\chi^2_{stat}$ and $\chi^2_{pull}$).

%%%%%%%%%%%%%%%%%%%%%%%%%%%%%%%%%%%%%%%%%%%%%%%%%%%%%%%%%%%
% Table for no charge id
%%%%%%%%%%%%%%%%%%%%%%%%%%%%%%%%%%%%%%%%%%%%%%%%%%%%%%%%%%%

%\begin{table}
%\TABLE[!h]{\centerline{
%\begin{center}
%\TABULAR
%\begin{tabular}{| c || c | c | c || c| c | c |} \hline
%\\
%$\sin^2 2 \theta_{13}$
%        & $(\chi^2_{m})^{stat}_{noID}$ &  $(\chi^2_{el})^{stat}_{noID}$ & $(\chi^2_{tot})^{stat}_{noID}$ & $(\chi^2_{m})^{pull}_{noID}$ & $(\chi^2_{el})^{pull}_{noID}$  & $(\chi^2_{tot})^{pull}_{noID}$
%                \\
                %&& \\
%       \hline \hline
%    0.0 &  20.1 \  7.1 \   27.2  &  0.0 & 27.2 & 25.0 & 0.0 & 25.0  \\ \hline
%    0.02  &  16.6 \ 7.4 \ 24.0 & 2.5 & 26.5 & 22.3 & 1.9 & 24.2   \\ \hline
%    0.04  &  13.6 \ 8.0 \ 21.6 & 6.3 & 27.9 & 20.0 & 4.7 & 24.7   \\ \hline
%    0.10  &  5.5 \ 8.7 \ 14.2 & 12.3 & 26.5 & 13.3 & 8.7 & 22.0
%     \\
%    \hline \hline
%    \end{tabular}
% \caption{Values of fixed parameter $\chi^2$ with and without pull for a Liquid Argon detector (1 Mt yr) {\bf{with no charge identification capability}}. The contributions from muon and electron events are also shown separately.}
%   \label{fixnoID}
%   \end{center}
%  \end{table}

\begin{table}
%\TABLE[!h]{\centerline{
\begin{center}
%\TABULAR
\begin{tabular}{| c || c | c | c || c| c | c |} \hline
%\\
$\sin^2 2 \theta_{13}$
        & $(\chi^2_{m})^{stat}$ &  $(\chi^2_{el})^{stat}$ & $(\chi^2_{tot})^{stat}$ & $(\chi^2_{m})^{pull}$ & $(\chi^2_{el})^{pull}$  & $(\chi^2_{tot})^{pull}$
        \\
%&& \\
        \hline \hline
%        0.0 &  20.1 \ 7.1 \ 27.2  &  0.0 & 27.2 & 25.1 & 0.0 & 25.1  \\ \hline
%        0.02  &  16.7 \ 10.1 \ 26.8 & 4.3 & 31.1 & 24.1 & 3.0 & 27.1   \\ \hline
        0.04  &  28.5 & 9.4 & 37.9 & 24.5 & 7.0 & 31.5   \\ \hline
        0.10  &  31.7 & 14.7 & 46.4 & 25.1 & 9.3 & 34.4
          \\
          \hline \hline
\end{tabular}
\caption{Values of fixed parameter $\chi^2$ without and with pull
for a Liquid Argon detector (1 Mt yr) with charge sensitivity. The contributions from muon
and electron events are also shown separately.}
%For \HKd, the values are $\chi^2_{\mu + {\bar{\mu}}} + \chi^2_{e + {\bar{e}}}$.
%For \INOd, the values are for $\chi^2_{\mu} + \chi^2_{\bar{\mu}}$.}
\label{fix}
\end{center}
\end{table}
%%%%%%%%%%%%%%%%%%%%%%%%%%%%%%%%%%%%%%%%%%%%%%%%%%%%%%%%%%55

%%%%%%%%%%%%%%%%%%%%%%%%%%%%%%%%%%%%%%%%%%%%%%%%%%%%%%%%%%%
%%%%%%% HKINO
%%%%%%%%%%%%%%%%%%%%%%%%%%%%%%%%%%%%%%%%%%%%
\begin{table}
%\TABLE[!h]{\centerline{
\begin{center}
%\TABULAR
\begin{tabular}{ |c || c | c | c | c | } \hline
%\\
$\sin^2 2 \theta_{13}$ & $(\chi^2_{m})^{pull}_{min}$ & $(\chi^2_{el})^{pull}_{min}$ & $(\chi^2_{tot})^{pull}_{min}$
        \\
%&& \\
        \hline \hline
%        0.0 &  0.0  & 0 & 0   \\ \hline
%        0.02  & 7.0 & 4.0 & 3.0    \\ \hline
        0.04  &  6.1 & 7.7 & 13.8  \\ \hline
        0.10  &  16.2 & 11.3 & 27.5 
          \\
          \hline \hline
\end{tabular}
\caption{Values of total marginalized $\chi^2$ with pull and priors,
for a Liquid Argon detector (1 Mt yr) with charge sensitivity. Also shown separately are the 
contributions 
of muon and electron events to the total $\chi^2$.}
%For \HKd, the values are $\chi^2_{\mu + {\bar{\mu}}} + \chi^2_{e + {\bar{e}}}$.
%For \INOd, the values are for $\chi^2_{\mu} + \chi^2_{\bar{\mu}}$.}
\label{margin}
\end{center}
\end{table}
%%%%%%%%%%%%%%%%%%%%%%%%%%%%%%%%%%%%%%%%%%%%%%%%%%%%%%%%%%55

\begin{table}
%\TABLE[!h]{\centerline{
\begin{center}
%\TABULAR
\begin{tabular}{ |c || c | c | c | c | } \hline
%\\
$\sin^2 2 \theta_{13}$ & $(\chi^2_{m})^{pull}_{min}$ & $(\chi^2_{el})^{pull}_{min}$ & $(\chi^2_{tot})^{pull}_{min}$
        \\
%&& \\
        \hline \hline
%        0.0 &  0.0  & 0 & 0   \\ \hline
%        0.02  & 7.0 & 4.0 & 3.0    \\ \hline
        0.04  & 1.3  & 4.9  & 6.2  \\ \hline
        0.10  & 3.5  & 9.2 & 12.7 
          \\
          \hline \hline
\end{tabular}
\caption{Values of total marginalized $\chi^2$ with pull and priors,
for a Liquid Argon detector (1 Mt yr) without charge sensitivity. Also shown separately are the
contributions
of muon and electron events to the total $\chi^2$.}
%For \HKd, the values are $\chi^2_{\mu + {\bar{\mu}}} + \chi^2_{e + {\bar{e}}}$.
%For \INOd, the values are for $\chi^2_{\mu} + \chi^2_{\bar{\mu}}$.}
\label{noID}
\end{center}
\end{table}

%%%%%%%%%%%%%%%%

%%%%%%%%%%%%%%%%%%%%

%%%%%%%%%%%%%%%%

	  %%%%%%%%%%%%%%%%%%%%

%--------------------------------------------------------------%
%               Figure 
%--------------------------------------------------------------%
%\begin{figure}
%\centerline{\psfig{figure=PmumuNH-IHvsE_L10000_m2.5.eps,height=18cm,width=20cm}}
%\caption{ Hierarchy difference for vacuum oscillations, for
%OMSD approximation and for the general case when both matter effects
%and $\Delta_{21}$ induced effects are present for various values of
%$\theta_{13}$. }
%\label{fig:hdiff}
%\end{figure}

%%%%%%%%%%%%%%%%

%%%%%%%%%%%%%%%%%%%%

%%%%%%%%%%%%%%%%%%%%

%%%%%%%%%%%%%%%%%%%%%%%%
%

\section{Summary and Conclusions}

From the results described in the previous section, we note the following features: 
\begin{itemize}

%\item 
%Table \ref{fixnoID} gives the values of fixed parameter $\chi^2$ for a 
%liquid Argon detector with no charge identification power. 
%We define 
%$(\chi^2_{tot})_{noID} = (\chi^2_{
%m})_{noID}+(\chi^2_{el})_{noID}$.

\item
Table \ref{fix} and Figure \ref{fig1} give the values of 
fixed-parameter $\chi^2$ for muon events with charge identification 
capacity and electron events with partial charge identification capacity 
as discussed in the previous section. 
We see that the values 
of fixed-parameter $(\chi^2_{tot})^{pull}$ are uniformly high 
($> 5\sigma$ for both input values of
$\theta_{13}$) for a Liquid Argon detector with 1 Mt yr exposure. 
Since earth matter effects are proportional to $\sin^2 2\theta_{13}$,
the $\chi^2$ is expected to vary linearly with $\sin^2 2\theta_{13}$.

\item
From Figure \ref{fig2}, we observe that the dependence of the total $\chi^2$ on the charge identification
capability for electron events is not drastic, since the contribution
from the electron events in this energy range represent only about
one-fifth of the total hierarchy
sensitivity of a Liquid Argon detector.
%Just for a check, we computed the $\chi^2$ for a Liquid Argon 
%detector with no charge identification at all (even for muons).
%These $\chi^2$ values are listed in Table \ref{noID}.

\item
%Table \ref{margin} and Figure \ref{fig3} depict the values of the total $\chi^2$ sensitivity
%to the neutrino mass hierarchy with marginalization over the neutrino parameters
%for a Liquid Argon detector with charge sensitivity.
%Table \ref{noID} gives the values of the marginalized hierarchy sensitivity for a Liquid Argon
%detector with no charge sensitivity.
Comparing the marginalized results with and without charge sensitivity
in Table \ref{margin} and Table \ref{noID}, we can see that $\chi^2$
in general increases after incorporation of charge identification
power of the detector. This is expected since the hierarchy sensitivity
comes mainly from the matter effect which is different for neutrinos
and antineutrinos. Hence a detector with charge sensitivity is better
capable of probing this.
Note that the muon $\chi^2$ shows a sharper rise with the inclusion
of charge sensitivity than the electron $\chi^2$, since 
charge identification with high precision is possible
over the entire energy range in the case of muons.

%Here, it is to be noted that the $\chi^2$ values reduce to zero for $\theta_{13}=0$ for muon events as well as electron events.
%This is because the marginalization over different values of $\vert \Delta m_{31}^2 \vert$ in $N^{th}$ 
%effectively washes out the hierarchy difference in $\pmumu$ for $\theta_{13}=0$. It can be verified that the difference
%$\vert \pmumu(NH) - \pmumu(IH) \vert$ for the hierarchy-sensitive $\theta_{13}$-independent term discussed above 
%becomes nearly zero if we choose $\Delta m_{31}^2 = -2.377 \times 10^{-3}$ eV$^2$ in $\pmumu(IH)$. 
%This causes a sharp drop in the $(\chi^2_{tot})^{pull}_{min}$ curve for small $\theta_{13}$, evident
%in Figure \ref{fig3}.
%

\item
From our study, a $ > 4\sigma$ sensitivity to the neutrino mass hierarchy 
is predicted for a magnetized Liquid Argon detector
with 1 Mt yr exposure for $\sin^2 2\theta_{13} > 0.05$ or $\theta_{13} > 6^\circ$. 
We redid our calculation for smaller exposure
of 333 kT yr, which can be achieved by   
a smaller Liquid Argon detector (30 kT, 11 years) with charge sensitivity. 
In this case, we obtain the marginalized total $\chi^2
= 4.9$, which is  $ > 2\sigma$ hierarchy sensitivity, for  $\sin^2 2\theta_{13} = 0.04$,
and  the marginalized total
$\chi^2 = 9.8$, which is  $ > 3\sigma$ hierarchy sensitivity, 
for  $\sin^2 2\theta_{13} = 0.1$.

\item
For a detector without charge sensitivity, a $2.5\sigma$ marginalized 
hierarchy sensitivity for $\sin^2 2\theta_{13} = 0.04$ 
is possible with 1 Mt yr exposure.

\end{itemize}

\section{Acknowledgments}
R.G. and S.G. acknowledge support from the XIth plan
neutrino project of Harish-Chandra Research Institute, Allahabad. S.G. and
S.U.S acknowledge partial support from the BRNS project. 
P.G, R.G. and S.U.S acknowledge hospitality 
of the theory group of Physical Research Laboratory, Ahmedabad
during the
finishing stage of this work.

%%%%%%%%%%%%%%%%%%%%%%%%%%%%%%%%%%%%%%%%%

\bibliographystyle{apsrevwinter}

\bibliography{myrefjan_08}

\end{document}